\documentclass[%
 reprint,
 amsmath,amssymb,
 aps,
 floatfix,
]{revtex4-1}

\usepackage{graphicx}
\graphicspath{{images/}}
\usepackage{dcolumn}
\usepackage{bm}

\begin{document}

\preprint{APS/123-QED}

\title{Relaxation to the equilibrium in the hard disk dynamics}

\author{Liliia Ziganurova $^{1,2}$}
\author{Lev N. Shchur $^{1,2}$}%
\affiliation{%
 $^1$ Science Center in Chernogolovka, 142432 Chernogolovka, Russia \\
 $^2$ National Research University Higher School of Economics,  101000 Moscow, Russia
 }%

\date{\today}

\begin{abstract}
We examine the question of the criteria of the relaxation to the equilibrium in the hard disk dynamics. In the Event-Chain Monte Carlo, we check the displacement distributions which follows to the exponential law. 
\end{abstract}

\maketitle


We restrict our current analysis to the case of the hard disks. 
In molecular dynamics simulations, it is important to know the relaxation time from the initial state to equilibrium. Which quantities of the equilibrium can give information that system reach equilibrium? In conventional Event Driven Molecular Dynamics it is the relaxation time to the Maxwell distribution which can be used as the estimate. In the case of the Event-Chain Monte Carlo~\cite{EC09} the velocities are not defined. Instead, we can use distribution of the displacements. It is known from the computer simulations~\cite{Alder65} that displacements $\delta r$  follow to the exponential law in the equilibrium

\begin{equation}
P(\Delta r) = \frac 1\lambda \exp{\left(-\frac{\Delta r}{\lambda}\right)},
\label{exp-distr}
\end{equation}
and theoretical arguments based on the kinetic theory support the findings~\cite{Turnbull70}.

We report the preliminary results for the system of a small number of disks $N=16$ in the box of the linear size $L=1.0$. 

\begin{figure}
\center{\includegraphics[width=1\linewidth]{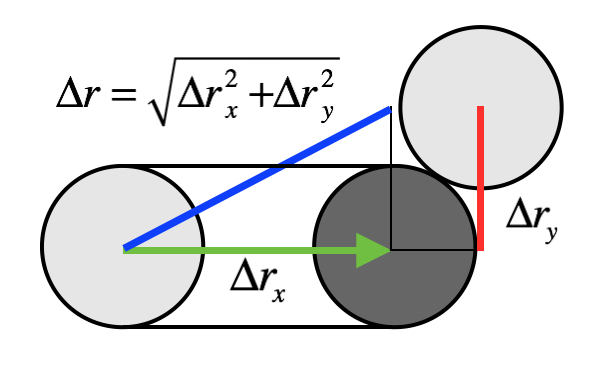}}
\caption{Definition of the displacements for the Event Chain Monte Carlo.}
\label{disp-ECMC}
\end{figure}

Definition of displacements is given in the Figure~\ref{disp-ECMC}. The real shifts $\Delta r_x$ are performed along the $x$-axes, and the formal shifts $\Delta r_y$ are defined in the perpendicular direction, as the difference of the projections of the centrum of colliding disks on the $y$-axes.  We also define the total displacement  $\Delta r$ as shown in the Figure~\ref{disp-ECMC}.

\begin{figure}
\center \begin{minipage}[h]{1\linewidth}
\center{\includegraphics[width=1\linewidth]{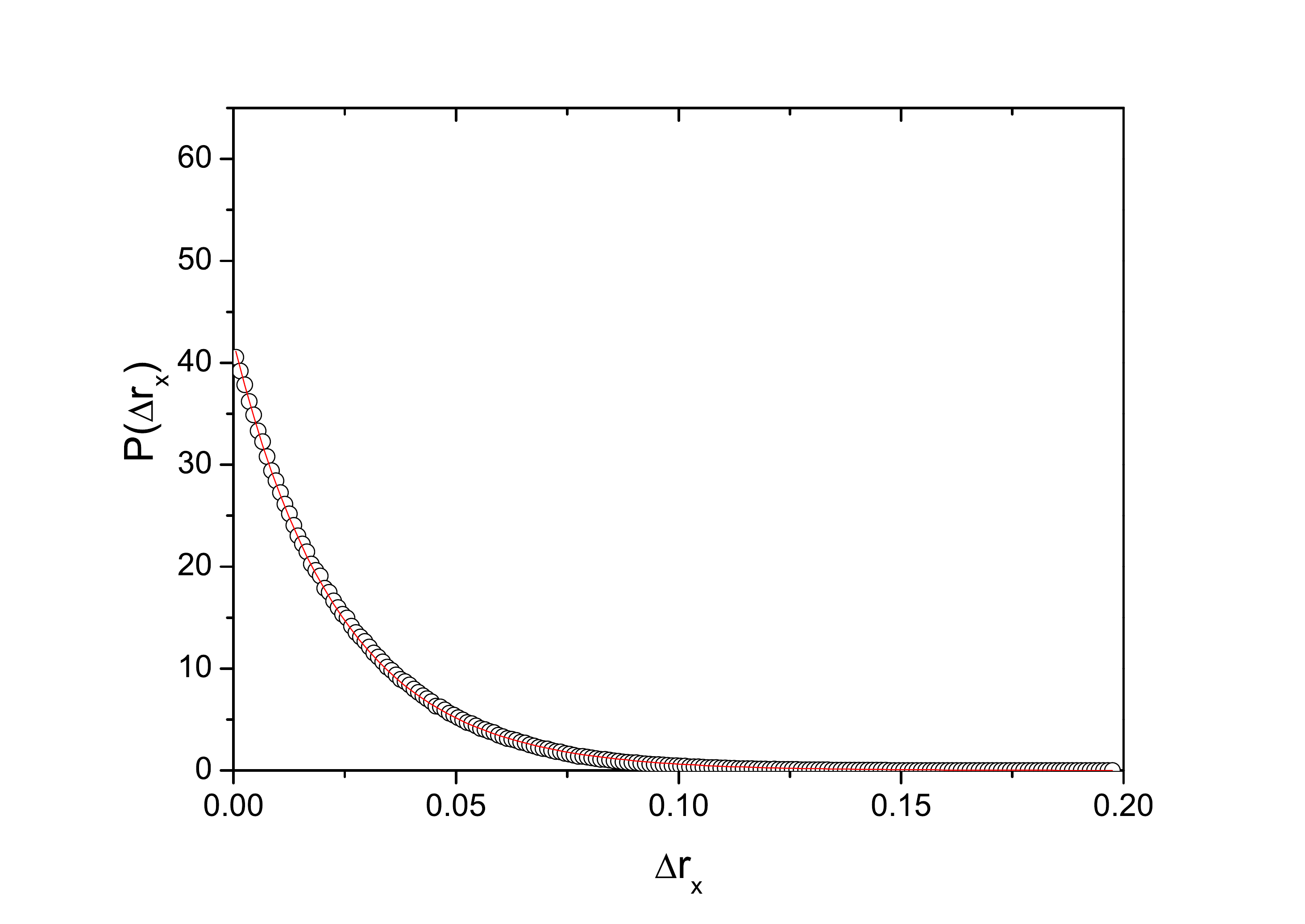}} a) $\eta = 0.68$ \\
\end{minipage}
\vfill
\begin{minipage}[h]{1\linewidth}
\center{\includegraphics[width=1\linewidth]{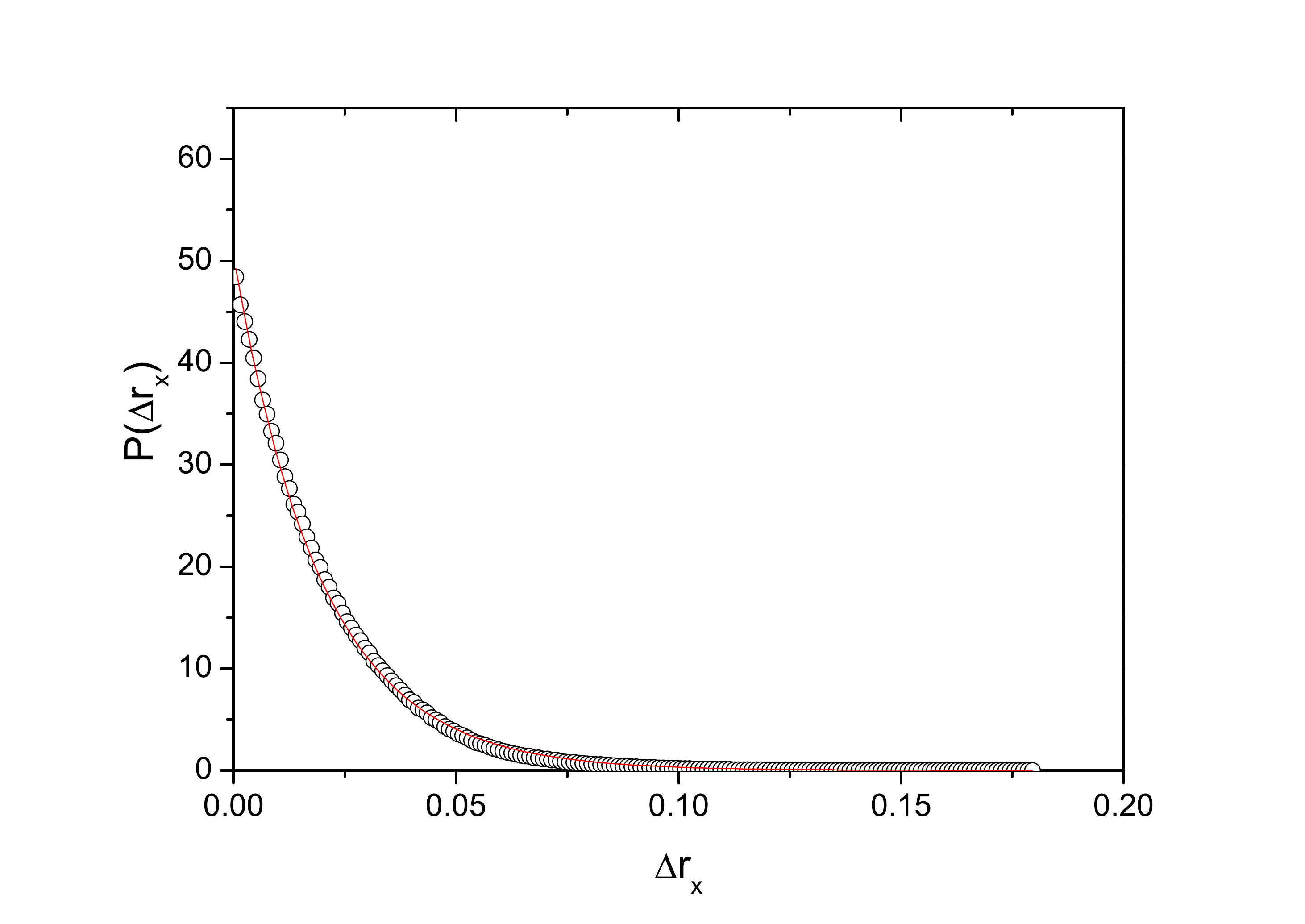}} \\b) $\eta = 0.71$
\end{minipage}
\vfill
\begin{minipage}[h]{1\linewidth}
\center{\includegraphics[width=1\linewidth]{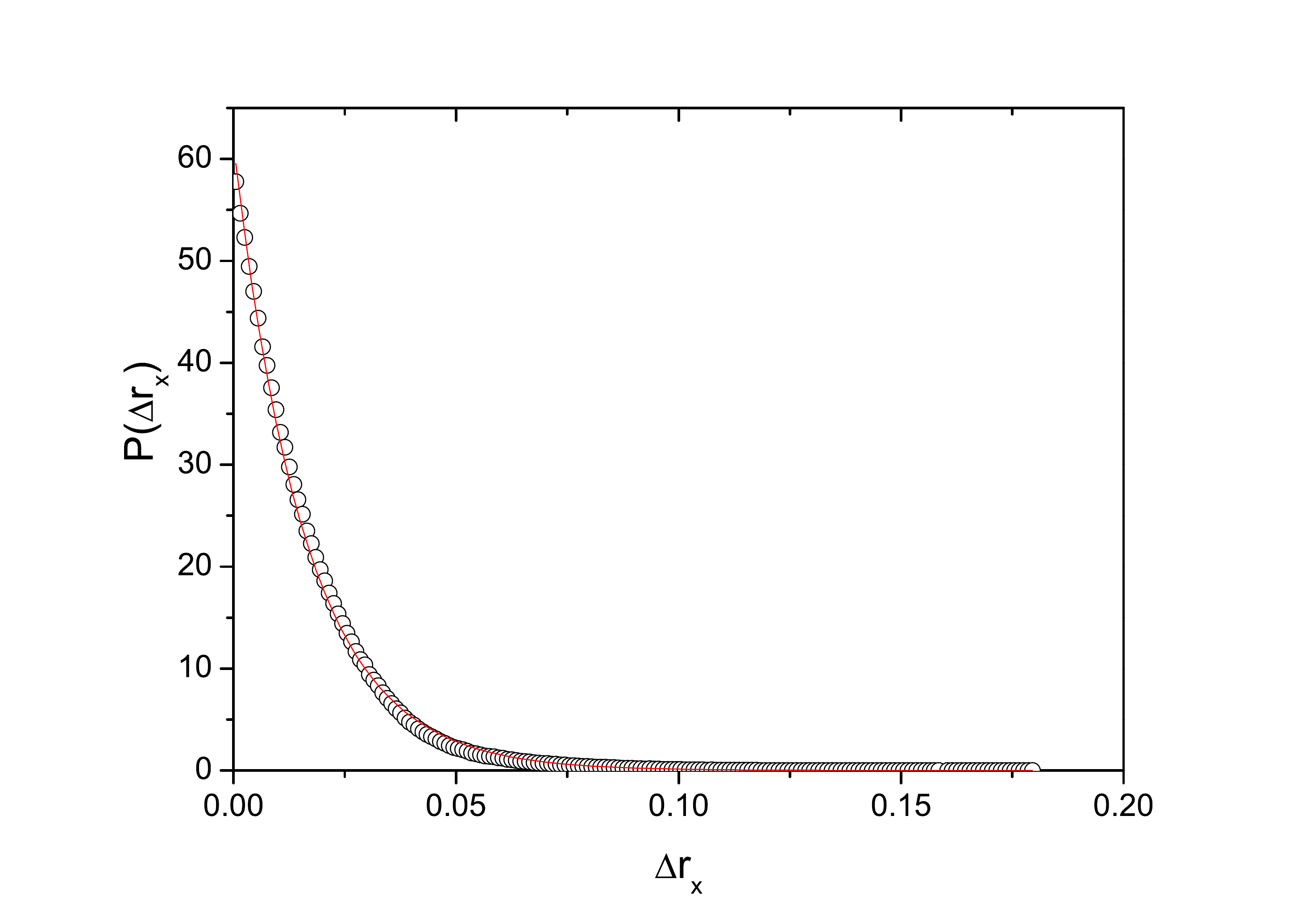}} c) $\eta = 0.73$\\
\end{minipage}
\vfill
\caption{Distribution of the displacements $\Delta r_x$ simulated using ECMC for the system of $N=16$ hard disks, for three values of densities $\eta$. Number of chains $M=10^6$.}
\label{distr-ECMC}
\end{figure}

We plot distribution of the displacements $\Delta r_x$ in the direction of the moves for the ECMC in the Figure~\ref{distr-ECMC} for three values of densities, $\eta=068, 0.71, 0.73$. The corresponding values of the mean free path are given in the Table~\ref{approx_of_Delta_x}. Value of $\lambda$ is defined with the density $\eta$ as $\lambda=\sqrt{L^2/\pi N} - \sigma$, the free volume for the particle. Estimation of the value $\lambda_{fit}$ from the fit is close to the free-path value $\lambda$ for the low density, as can be expected from the definition of $\lambda$, and diverges for the large densities $\eta$.

\begin{table}[h]
\label{approx_of_Delta_x}
\caption{Approximation of the distributions of the displacements $\Delta r_x$ pictured in the Figure~\ref{distr-ECMC}. Density $\eta$ defines the disk radius $\sigma$ and the free path value $\lambda$. The $\lambda_{fit}$ is the value of the fit to expression~(\ref{exp-distr}) and $\chi^2$ is the valued of the $chi^2$.}
\begin{center} 
\begin{tabular}{|l|l|l|l|l|}
\hline
$\eta$     &    $\sigma$     &        $\lambda$   & $\lambda_{fit}$   &        $\chi^2$    \\
\hline
0.68         &    0.116311    &     0.02474  &  0.02382(4)    &    0.037     \\
\hline
0.71         &    0.118849    & 0.02220   &    0.01980(7)    &    0.138    \\
\hline
0.73      &     0.120511      & 0.02056  &    0.01628(6)        &    0.226       \\\hline
\end{tabular}
\end{center}
\end{table}

In the ECMC, the direction of the moves alternates from $x$-axes to the $y$-axes, therefore all real shifts of the disks follows to the expected distribution~(\ref{exp-distr}) and provides the equilibrium state of the system. In a way, this is an additional argument for the validity of the Event-Chain Monte Carlo  approach for simulation of hard disks (and spheres). 

The work is supported by the grant 17-07-01537 from Russian Foundation for Basic Research.

\end{document}